\documentstyle[12pt]{article}
\def\thesection {\arabic{section}}

\def\theequation{\thesection.\arabic{equation}}

\begin{document}

\centerline{IBR   preprint   TH--97--S037, February   15,  1997}
\centerline{PACS  03.20+i}

\vspace*{2cm}
\begin{center}
{\Large \bf A CLASSICAL ISODUAL THEORY OF ANTIMATTER}
\end{center}

\vskip 1cm
\begin{center}
{\it {\bf  Ruggero Maria  Santilli}\\  Institute for  Basic Research\\
P. O. Box  1577, Palm Harbor,  FL 34682,  U.S.A.}\\ {\sl ibr@gte.net,
http://home1.gte.net/ibr}
\end{center}

\vskip 1.5cm
\begin{abstract}
An inspection of  the  contemporary physics literature  reveals  that,
while matter is  treated  at   all levels  of study,  from   Newtonian
mechanics to quantum field theory, antimatter is solely treated at the
level  of second quantization.  For    the purpose of initiating   the
restoration  of  full equivalence  in   the  treatments of matter  and
antimatter  in  due  time,  in  this  paper  we  present  a  classical
representation of  antimatter which begins  at the primitive Newtonian
level with expected images at all subsequent levels. By recalling that
charge    conjugation     of    particles     into    antiparticles is
anti--automorphic, the proposed theory of antimatter is based on a new
map, called {\it isoduality}, which  is  also anti--automorphic, yet
it  is
applicable beginning  at the classical level and  then persists at the
quantum level. As part of our study, we present novel anti--isomorphic
{\it isodual images of the  Galilean, special and general relativities}
and
show the compatibility of  their representation of antimatter with all
available classical  experimental  knowledge,  that on electromagnetic
interactions.  We  also  identify the prediction    of antigravity for
antimatter in the field  of matter (or  vice--versa) without any claim
on its validity, and defer its  resolution to specific experiments. To
avoid a  prohibitive length, the paper is  restricted to the classical
treatment which had not  been sufficiently treated until  now. Studies
on operator profiles, such as the equivalence of isoduality and charge
conjugation and  the implication of   the  isodual theory in  particle
physics, are conducted n a separate paper.
\end{abstract}

\section{INTRODUCTION}
\setcounter{equation}{0}        After being conjectured by A. Schuster
in 1898, antimatter was
predicted  by P.  A.  M. Dirac~[1]  in the  late  1920's  in the  {\it
negative--energy solutions} of his celebrated equation.  Dirac himself
soon discovered that particles with negative--energy  do not behave in
a physical way  and,  for this  reason,  he  submitted  his celebrated
``hole theory'', which subsequently restricted the study of antimatter
to the level   of  second  quantization (for   historical  aspects  on
antimatter see, e.g., Ref.~[2]).

        The  above occurrence created  an  imbalance in the physics of
this century because matter is described at all  levels of study, from
Newtonian    mechanics to quantum  field  theory,  while antimatter is
solely treated at the level of second quantization.

        To initiate the study for the future removal of this imbalance
in due time, in this paper we present a theory of antimatter which has
been conceived to begin  at the purely  classical Newtonian level, and
then to admit corresponding images at all subsequent levels of study.

        Our guiding principle is to identify a map which possesses the
main mathematical structure   of  charge   conjugation, yet it      is
applicable at all levels, and not solely at the operator level.

        The main characteristic  of charge conjugation is
that of being  {\it  anti--auto\-mor\-phic} (or, more generally,  {\it
anti--iso\-mor\-phic}).  After studying  a number of possibilities,  a
map which is anti--isomorphic and  applicable at all levels of  study,
is the following {\it isodual map} of any given quantity $Q$

\begin{equation}
  Q\;\;\;\rightarrow\;\;\; Q^d\;\; =\;\; - Q^{\dagger} .
%(1.1)
\label{eq:one-1}\end{equation}
which,  for consistency,  must be   applied  to  the  totality of  the
mathematical structure of the conventional theory of matter, including
numbers, fields, spaces, geometries,  algebras, etc. This results in a
new mathematics,  called {\it isodual  mathematics},  which is at  the
foundation  of the classical    isodual theory of  antimatter of  this
paper.

       Since  the isodual  mathematics is virtually  unknown, we shall
review  and expand it in  Sect.~2.  In Sect.~3  we shall then present,
apparently for the first   time, the classical {\it  isodual Galilean,
special and general relativities}  and show that their  representation
of  antimatter  is    indeed compatible   with  available    classical
experimental data, those of electromagnetic nature. In the Appendix we
outline the  classical  {\it   isodual  Lagrangian  and    Hamiltonian
mechanics}. The  operator version of  the isodual theory of antimatter
is studied  in  a separate  papers which  also prove   the equivalence
between isoduality and charge conjugation.

        The  rather limited  existing literature in  isoduality is the
following.  The  isodual  map~(\ref{eq:one-1}) was  first  proposed by
Santilli  in Ref.s~[3] of 1985  and then  remained ignored for several
years.  More recently, the    {\it isodual numbers}  characterized  by
map~(\ref{eq:one-1})   have been  studied   in  paper~[4].  The  first
hypothesis   on the isodual  theory   of antimatter  appeared for  the
operator  version in Ref.~[5]  of 1993 which  also contains an initial
study of  the  equivalence between isoduality and  charge conjugation.
The  fundamental  notion of this    study,  the  {\it  isodual
Poincar\`{e} symmetry}, from which  the entirety of the (relativistic)
analysis can  be uniquely  derived,   was submitted in
Ref.~[6] of  1993 also at  the  operator level.

                                                        The prediction 
of the
isodual   theory that antimatter in   the  field of matter experiences
antigravity was first  submitted in Ref.~[7] of  1994. An experiment for
the
measure of the gravity of elementary antiparticles in the gfravitational
field of Earth was also proposed in Ref.~[7]. It essentially consists in
commparative measures of the gravity of collimated, {\it low energy}
beams
of positrons and electrons in {\it horizontal} flight on a tube
with sufficienr vacuum
as well as protection from stray fields and of sufficient length to
permit
a definite result, e.g., the view by the naked eye of the displacements
due to gravity of the positron and electron beams on
a scientillator at the end of the flight.

                         The      {\it isodual
differential   calculus},  which    is  fundamental  for the   correct
formulation of dynamical  equations  all the  way  to those in  curved
spaces, was  studied only recently in  Ref.~[8]. A review  of the {\it
operator} profile up to 1995 is available in monograph~[9].

        This paper is  the classical   counterpart  of the   companion
paper~[10]  in  which we  study the operator  profile, with particular
reference to the equivalence between isoduality and charge conjugation
and  the  predictions  of the  new   theory  in  particle physics.  The
additional adjoining  paper~[11]   presents a general outline  of  the
isodualities of the broader {\it  isotopic,
genotopic and  hyperstructural}  formulations
which are under study for antimatter in {\it interior} conditions
(such as for the structure of an antimatter star) and they
are not considered in this paper for brevity.

        An important independent  contribution  in the field  has
been made  by the   experimentalist   A.P.~Mills, Jr.~[12],   who  has
confirmed the apparent feasibility with current technology of the test
of  the gravity   of antiparticles proposed    in ref.~[7] via the
use of eletrons and positrons with energy of the order of 
milli-eV in horizontal flight in a vacuum
tube of
approximately 100 m length with a diameter and design suitable to
reduce stray fields and patch effects at its center down to acceptable
levels.

        Additional
contributions   have been made    by: J. V. Kadeisvili  on  the {\it
isodual
functional analysis} and {\it isodual  Lie theory}~[13];
Lohmus, Paal, Sorgsepp, Sourlas, Tsagas~[14]; and others.

        Theoretical and experimental studies on  the isodual theory of
antimatter  were conducted at  the    {\it International Workshop   on
Antimatter Gravity  and  Anti--Hydrogen  Atom Spectroscopy}, held   in
Sepino, Italy, in May 1996 (see the Proceedings~[15]).

\section{RUDIMENTS OF ISODUAL MATHEMATICS}
\setcounter{equation}{0}
\subsection{Isodual  units, numbers, and  fields.}  Let $F\; =\;
F(a,+,\times )$ be a conventional field  of real numbers $R(n,+,\times
)$,  complex  numbers  $ C(c,+,\times   )  $ or  quaternionic  numbers
$Q(q,+,\times )$ with  the familiar additive unit $0$,  multiplicative
unit $I$, elements $a\;=\; n, c, q$, sum $a_1 + a_2$, $a + 0\;=\;0 + a
\;=\; a$,  and  multiplication $a_1\times a_2  \;=\; a_1a_2$, $a\times
I\;= \; I\times a \;= \; a,\;\;\forall a, \; a_1, a_2 \in F$.

        The {\it  isodual fields},  first introduced  in  ref.~[3] and
then  studied  in  details in   ref.~[4],  are the  image   $F^d \;=\;
F^d(a^d,+^d,\times^d)$ of $F(a,+,\times)$ characterized by the isodual
map of the unit

\begin{equation}
   I\;\; \rightarrow I^d \;=\; - I^\dagger \;=\; - I ,
%   (2.1)
\label{eq:two-1}\end{equation}
which implies: {\it isodual numbers}

\begin{equation}
a^d \;=\;  a^\dagger \times  I^d \;=\;  - a^\dagger  \times I  \;=\; -
a^\dagger ,
%    (2.2)
\label{eq:two-2}\end{equation}
where $^\dagger$ is the identity for real numbers $n^\dagger \;=\; n$,
complex conjugation $c^\dagger\;=\; \bar{c}$ for complex numbers c and
Hermitean  conjugation $q^\dagger$  for quaternions  $q^\dagger$; {\it
isodual sum}

\begin{equation}
              a_1^d +^d a_2^d \;=\; - ( a_1^\dagger + a_2^\dagger ) ;
% (2.3)
\label{eq:two-3}\end{equation}
and {\it isodual multiplication}

\begin{equation}
      a_1^d \times^d a_2^d \;=\; a_1^d \times I^d \times a_2^d = \;=\;
- a^\dagger_1 \times a^\dagger_2 ;
%(2.4)
\label{eq:two-4}\end{equation}
under which $I^d$ is the correct left and right unit of $F^d$,

\begin{equation}
I^d \times^d a^d \;=\;  a^d \times^d I^d \equiv  a^d ,\;\; \forall a^d
\in F^d ,
%  (2.5)
\label{eq:two-5}\end{equation}
in which case (only) $\hat{I}^d$ is called {\it isodual unit}.

We    have     in    this   way  the   {\it      isodual  real  field}
$R^d(n^d,+^d,\times^d)$ with {\it isodual real numbers}

\begin{equation}
n^d \;=\; - n^\dagger \times I \equiv - n ,\;\; n  \in R ,\;\; n^d \in
R^d;
%  (2.6)
\label{eq:two-6}\end{equation}
the  {\it isodual  complex   field} $C^d(c^d,+^d,\times^d)$ with  {\it
isodual complex numbers}

\begin{eqnarray}
\lefteqn{c^d \;=\; -  \bar{c} \;=\; - (  n_1 - i  \times n_2 ) \;=\; -
n_1 +  i \times n_2,} \nonumber  \\ & &\mbox{}n_1, n_2 \in  R , \;\;
c \in C, \;\; c^d \in C^d;
% (2.7)
\label{eq:two-7}\end{eqnarray}
and the  {\it isodual quaternionic field}   which is not  used in this
paper for brevity.

        Under  the above assumptions, $F^d(a^d,+^d,\times^d)$ verifies
all the  axioms  of a field~[loc.  cit.],  although $F^d$ and $F$  are
anti--isomorphic, as   desired.   This establishes that  the  field of
numbers can be equally defined  either with respect to the traditional
unit +1 or with respect to its negative image -1. The key point is the
preservation   of  the axiomatic    character of  the  latter  via the
isoduality of  the  multiplication. In other   words, the set  isodual
numbers $a^d$ with unit -1 and {\it  conventional} product {\it does
not}
constitute a field because $I^d\times a^d \;\not=\; a^d$.

It  is   also evident  that {\it  all   operations of numbers implying
multiplications must be subjected for consistency to isoduality}. This
implies the {\it isodual square root}

\begin{equation}
   a^{d\frac{1}{2}d}  \;=\;  -  \sqrt{-a^d},\;\;  a^{d\frac{1}{2}d}
\times^d a^{d\frac{1}{2}d} \;=\;  a^d , \;\;\; 1^{d\frac{1}{2}d} \;=\;
-i
;
% (2.8)
\label{eq:two-8}\end{equation}
the {\it isodual quotient}

\begin{equation}
a^d /^d b^d \;=\; - ( a^d / b^d ) \;=\;  - ( a^\dagger / b^\dagger
=  ) \;=\; c^d ,\;\;
b^d \times^d c^d \;=\; a^d ;
% (2.9)
\label{eq:two-9}\end{equation}
and so on.

        A property of isodual fields  of fundamental relevance for our
characterization   of   antimatter     is  that   {\it    they    have
negative--definite norm}, called {\it isodual norm}~[3,4]

\begin{equation}
| a^d  |^d \;=\; |   a^\dagger | \times   I^d  \;=\; - (  a  a^\dagger
)^{1/2}\; < \; 0 .
%     (2.10)
\label{eq:two-10}\end{equation}
where $|  ...  |$  denotes  the conventional   norm. For  isodual real
numbers $n^d$ we therefore have the isodual isonorm

\begin{equation}
| n^d |^d \;=\; - | n | < 0 ,
% (2.11)
\label{eq:two-11}\end{equation}
and for isodual complex numbers we have

\begin{equation}
| c^d |^d  \;=\; - |\bar{c}|  \;=\; - (  c  \bar{c} )^{1/2} \;=\;  - (
n_1^2 + n_2^2 )^{1/2} .
% (2.12)
\label{eq:two-12}\end{equation}

\vskip  0.5cm {\bf  Lemma   2.1:}    {\sl All quantities  which    are
positive--definite when  referred to  fields  (such as  mass,  energy,
angular momentum,    density,    temperature,  time,  etc.)     became
negative--definite when referred to isodual fields.}

\vskip  0.5cm  As  recalled  in   Sect.  1, antiparticles   have  been
discovered  in the    {\it   negative--energy solutions}   of  Dirac's
equation~[1] and they were  originally thought to evolve {\it backward
in time}   (Stueckelberg,  and others, see~[2]).    The possibility of
representing  antimatter  and  antiparticles  via  isodual  methods is
therefore visible already from these introductory notions.

        The main novelty     is that the conventional  treatment    of
negative--definite energy and time was (and still  is) referred to the
conventional  contemporary  unit  +1,   which  leads to   a  number of
contradictions  in    the  physical behavior of    antiparticles whose
solution forced the transition to second quantization.

        By comparison, {\it the negative-definite physical quantities
of  isodual methods are referred to  a negative--definite unit} $I^d <
0$.  As  we   shall see, this   implies  a  mathematical  and physical
equivalence   between {\it  positive--definite quantities referred  to
positive--definite         units,     characterizing  matter,      and
negative--definite quantities  referred  to negative--definite  units,
characterizing   antimatter}. These  foundations  then  permit a novel
characterization of antimatter beginning at the {\it Newtonian} level,
and then persisting at all subsequent levels.

\vskip   0.5cm   {\bf  Definition  2.1:}{\sl   A   quantity  is called
isoselfdual when it is invariant under isoduality.}

\vskip 0.5cm The above notion is particularly important for this paper
because  it introduces a new    invariance, {\it the invariance  under
isoduality}.  We  shall encounter  several  isoselfdual quantities. At
this introductory  stage we indicate that {\it  the imaginary number i
is isoselfdual},

\begin{equation}
i^d \;=\; - i^\dagger \;=\; - \bar{i} \;=\; - ( - i ) \;=\; i .
% (2.13)
\label{eq:two-13}\end{equation}

This  property permits to  understand better the isoduality of complex
numbers which can be written explicitly~[4]

\begin{equation}
c^d \;=\; ( n_1 + i\times  n_2 )^d \;=\;  n_1^d + i^d  \times^d n_2^d
\;=\; -
  n_1 + i \times n_2 \;=\; -\bar{c} .
%     (2.14)
\label{eq:two-14}\end{equation}

        We assume the  reader is aware of  the emergence here of  {\it
basically new   numbers, those with  a  negative unit}, which  have no
connection with  ordinary negative  numbers   and which are   the true
foundations of the proposed isodual theory of antimatter.

\vskip 0.5cm

\subsection{Isodual  functional analysis.} All conventional and  special
functions and transforms, as well as functional analysis at large must
be  subjected to isoduality   for  consistent applications of  isodual
theories,    resulting in a simple,   yet  unique and significant {\it
isodual   functional  analysis},    whose  study   was  initiated   by
Kadeisvili~[13].

         We here mention the {\it isodual trigonometric functions}

\begin{equation}
\sin^d \theta^d \;=\; - \sin ( - \theta ) ,\;\; \cos^d= \theta^d \;=\;
 - \cos ( - \theta ) ,
%  (2.15)
\label{eq:two-15}\end{equation}
with related basic property

\begin{equation}
\cos^{d\: 2d} \theta^d + \sin^{d\: 2d} \theta^d \;=\; 1^d \;=\; - 1 ,
% (2.16)
\label{eq:two-16}\end{equation}
the {\it isodual hyperbolic functions}

\begin{equation}
\sinh^d w^d \;=\; - \sinh ( - w ) ,\; \cosh^d w^d \;=\;  - \cosh ( - w
) ,
%  (2.17)
\label{eq:two-17}\end{equation}
with related basic property

\begin{equation}
\cosh^{d\: 2d} w^d - \sinh^{d\: 2d} w^d \;=\; 1^d \;=\; -1 ,
%  (2.18)
\label{eq:two-18}\end{equation}
the {\it isodual logarithm}

\begin{equation}
\log^d n^d \;=\; - log ( - n) ,
% (2.19)
\label{eq:two-19}\end{equation}
etc. Interested readers can then easily construct the isodual image of
special functions, transforms, distributions, etc.

\vskip 0.5cm

\subsection{Isodual   differential  calculus.}   The   conventional
differential  calculus is indeed  dependent  on the assumed unit. This
property is not so transparent in the conventional formulation because
the  basic   unit   is  the trivial  number    +1,   thus having  null
differential.  However,  the dependence  of  the  unit emerges  rather
forceful under its generalization.

        The  {\it isodual differential  calculus}, first introduced in
ref.~[8], is characterized by the {\it isodual differentials}

\begin{equation}
d^d x^k \;=\; I^d\times dx^k \;=\; - dx^k, \;\; d^d x_k \;=\; - d x_k ,
%   (2.20)
\label{eq:two-20}\end{equation}
with corresponding {\it isodual derivatives}

\begin{equation}
\partial^d /\partial^d  x^k  \;=\; - \partial  / \partial   x^k , \;\;
\partial^d / \partial^d x_k \;=\; -\partial / \partial x ,
%  (2.21)
\label{eq:two-21}\end{equation}
and other isodual properties.

Note that {\it conventional differentials are isoselfdual}, i.e.,

\begin{equation}
( d x^k )^d \;=\; d^d x^{kd} \equiv d\: x^k ,
%(2.22)
\label{eq:two-22}\end{equation}
but {\it derivatives are not in general isoselfdual},

\begin{equation}
( \partial f(x) / \partial x^k )^d \;=\; \partial^d f^d /^d \partial^d
x^{kd} \;=\; - \partial f / \partial x^k.
%(2.23)
\label{eq:two-23}\end{equation}

        Other  properties can  be  easily derived  and shall be hereon
assumed.

\vskip 0.5cm

\subsection{Isodual Lie theory.} Let  {\bf L}  be an n--dimensional Lie
algebra with universal enveloping  associative algebra $\xi$({\bf L}),
$[\xi$({\bf L})$]^-  \approx$ {\bf L}, n-d-imensional unit  $I  \;=\;
diag. (1, 1,
..., 1)$ for   the regular representation,  ordered  set  of Hermitean
generators $X  \;=\; \;=\;  X^\dagger \;=\;  \{ X_k  \},$ conventional
associative product $X_i\times X_j$,  and familiar Lie's Theorems over
a field $F(a,+,\times ).$

        The {\it isodual Lie  theory} was first submitted  in ref.~[3]
and then studied  in Ref.~[9] as  well as by other authors~[13,14]. {\it
The   isodual universal    associative   algebra} $[\xi$({\bf
L})$]^d$    is
characterized   by   the  {\it  isodual    unit} $I^d$,   {\it isodual
generators} $ X^d \;=\; -X$, and isodual associative product

\begin{equation}
                 X_i^d \times^d X_j^d \;=\; - X_i \times X_j,
%       (2.24)
\label{eq:two-24}\end{equation}
with corresponding infinite--dimensional basis (isodual version of the
conventional Poincar\'{e}--Birkhoff--Witt theorem~[3])  characterizing
the {\it isodual exponentiation} of a generic quantity A

\begin{equation}
e^{d^A} \;=\; I^d + A^d /^d 1 !^d  + A^d \times^d  A^d /^d 2 !^d +
...  \;=\; - e^{A^\dagger} ,
%(2.25)
\label{eq:two-25}\end{equation}
where e is the conventional exponentiation

The attached  {\it isodual Lie  algebra} {\bf L}$^d \approx (\xi^d)^-$
over the isodual field $F^d(a^d,+^d,\times^d)$ is characterized by the
{\it isodual commutators}~[loc. cit.]

\begin{equation}
[  X_i^d  ,  X_j^d  ]^d \;=\;  -  [X_i, X_j] \;=\; C_{ij}^{kd}
\times^d X^d_k .
%    (2.26)
 \label{eq:two-26}\end{equation} with a classical realization given in
Appendix A.

Let    G  be   the conventional,   con\-nec\-ted,  n--dimensional  Lie
trans\-for\-ma\-tion group on  $S(x,$ $g,$ $F)$ admitting {\bf L} as  
the Lie
algebra in the neighborhood  of the  identity, with generators  $X_k$
and  parameters  $w \;=\;  \{w_k\}$.    The  {\it isodual  Lie  group}
$G^d$~[3] admitting the isodual Lie algebra  ${\bf L}^d$ in the
neighborhood
of the isodual identity Id is the n--dimensional group with generators
$X^d \;=\; \{-X_k\}$  and   parameters $w^d \;=\; \{-w_k\}$   over the
isodual field $F^d$ with generic element

\begin{equation}
     U^d(w^d)   \;=\;  e^{d^{i^d\times^d   w^d\times^d X^d}}   \;=
\; -   e^{i\times (-w)\times X}  \;=\;  -U(-w) .
%(2.27)
\label{eq:two-27}\end{equation}

        The {\it isodual symmetries} are  then defined accordingly via
the  use of the isodual groups  $G^d$ and they are anti--isomorphic to
the corresponding conventional  symmetries, as desired. For additional
details, one may consult Ref.~[9].

        In this paper we shall therefore use:
\begin{enumerate}
\item[{\bf   1)}]    {\bf  Conventional  Lie  symmetries,}   for   the
characterization of {\it matter}; and\\
\item[{\bf 2)}] {\bf Isodual Lie symmetries,} for the characterization
of {\it antimatter}.  \\
\end{enumerate}

\subsection{Isodual Euclidean  geometry.}  Conventional   (vector and)
metric spaces are defined over conventional fields. It is evident that
the isoduality  of fields requires,  for  consistency, a corresponding
isoduality  of  (vector   and)   metric  spaces.   The need   for  the
isodualities   of all   quantities acting on    a  metric space (e.g.,
conventional  and  special   functions and   transforms,  differential
calculus, etc.) becomes then evident.

        Let $S \;=\; S(x,g,R)$ be a conventional N--dimensional metric
space with local coordinates $x \;=\; \{x^k\}$, $k \;=\; 1, 2, ..., N$,
nowhere degenerate, sufficiently  smooth, real--valued  and  symmetric
metric $g(x, ...)$ and related invariant

\begin{equation}
x^2 \;=\; x^i g_{ij} x^j ,
% (2.28)
\label{eq:two-28}\end{equation}
over the reals R.

The {\it isodual spaces},   first introduced Ref.~[3], are the  spaces
$S^d(x^d,g^d,R^d)$ of  $S(x,g,R)$ with  {\it isodual coordinates} $x^d
\;=\; x\times I^d$, {\it isodual metric}

\begin{equation}
g^d(x^d, ...)  \;=\; - g^\dagger(-x, ...)  \;=\; - g(-x, ... ),
%  (2.29)
\label{eq:two-29}\end{equation}
and {\it isodual interval}

\begin{equation}
(x - y)^{d2\: d}  \;=\; [ ( x -  y)^{i d} \times^d
g_{ij}^d \times^d( x   - y )^{j d}  ]\times I^d
 = \; [ ( x -  y)^i \times
g_{ij}^d \times( x   - y )^{j d}  ]
\times I^d ,
%(2.30)
\label{eq:two-30}\end{equation}
defined over the  isodual field $R^d \;=\; R^d(n^d,+^d,\times^d)$ with
the same isodual isounit $I^d$.

The basic space of our analysis is the three--dimensional {\it isodual
Euclidean space},

\[
E^d(r^d,\delta^d,R^d) : r^d \;=\; \{ r^{kd} \} \;=\; \{-r^k\} \;=\; \{
-x, -y, -z \} ,
\]
\begin{equation}
\delta^d \;=\; -\delta \;=\; diag. ( -1, -1, -1  ) , \;\; I^d \;=\; -I
\;=\; diag. (-1, -1, -1) .
%  (2.31)
\label{eq:two-31}\end{equation}

The {\it   isodual Euclidean geometry}  is  then  the geometry of  the
isodual space  $E^d$ over  $R^d$ and  it is given  by a step--by--step
isoduality of all the various aspects of the conventional geometry.

We only mention for brevity the notion  of {\it isodual line} on $E^d$
over $R^d$  given by the  isodual image of  the conventional notion of
line on E  over R. As such, its  coordinates are isodual  numbers $x^d
\;=\; x\times 1^d$  with unit $1^d = \;=\;-1$.  By recalling that  the
norm on $R^d$ is negative--definite,  the {\it isodual distance} among
two points  on  an isodual line is  also  negative definite and it  is
given by    $D^d  \;=\; D\times   1^d  \;=\; -D$,    where $D$  is the
conventional distance.  Similar  isodualities  apply to  all remaining
notions, including the notions  of  parallel and intersecting  isodual
lines, the Euclidean axioms, etc. The following property is of evident
proof:

\vskip 0.5cm {\bf Lemma 2.2:} {\sl The  isoeuclidean geometry on $E^d$
over   $R^d$ is anti--isomorphic  to  the conventional geometry on $E$
over $R$.}

\vskip 0.5cm The {\it isodual  sphere}
        is the perfect sphere in $E^d$ over  $R^d$ and, as such,
        it has {\it negative radius},

\begin{equation}
R^{d2d} \;=\; [x^{d2d} + y^{d2d} + z^{d2d}]\times I^d \;.
%(2.32)
\label{eq:three-32}\end{equation}

A   similar characterization  holds   for other  isodual  shapes which
characterize the shape of antimatter in our isodual theory.

        The  group of  isometries   of $E^d$ over  $R^d$  is  the {\it
isodual euclidean group} studied in Ref.~[9].

\vskip 0.5cm

\subsection{Isodual  Minkowskian geometry.}  The {\it  isodual Minkowski
space}, first introduced in Ref.s~[3], is given by

\[
 M^d(x^d,\eta^d,R^d): x^d \;=\;  \{x^{\mu d}\}  \;=\; \{x^{\mu}\times
I^d\} \;=\;
\{-r , -c_o t\}\times I ,
\]
\begin{equation}
\eta^d  \;=\;  -\eta \;=\;  diag.  ( -1, -1,  -1,  +1 ),\;\; I^d \;=\;
diag. ( -1, -1, -1, -1 ) .
%  (2.33)
\label{eq:two-33}\end{equation}

The {\it isodual Minkowskian geometry}~[6]  is the geometry of isodual
spaces $M^d$   over  $R^d$. It   is also  characterized by   a  simple
isoduality  of the conventional  Minkowskian geometry and its explicit
presentation is omitted for brevity.

        We here merely mention the {\it isodual light cone}

\begin{equation}
x^{d\: 2\: d}  \;=\; (x^{\mu d}\times^d  \eta_{\mu\nu}^d\times^d x^{\nu
d})\times I^d
\;=\; ( - x\: x - y\: y - z\: z + t c_o^2 t ) \times ( - I ) \;=\; 0.
%   (2.34)
\label{eq:two-34}\end{equation}

As one   can  see,   the   above  cone  formally  coincides   with the
conventional light cone,  although the two  cones  belong to different
spaces.  The isodual light cone  is used in these  studies as {\it the
cone   of light  emitted by   antimatter   in  empty space   (exterior
problem)}.

        The  group of  isometries   of $M^d$ over  $R^d$  is  the {\it
isodual   Poincar\`{e}  symmetry}    $P^d(3.1)  \;=\;   L^d(3.1)\times
T^d(3.1)$~[6] and constitutes the fundamental symmetry of this paper.

\vskip 0.5cm

\subsection{Isodual Riemannian geometry.}   Consider a Riemannian space
$\Re(x,g,R)$ in $(3+1)$ dimensions with basic  unit $I \;=\; $$diag.$ $
(1,
$ $1,$ $ 1,$ $  1)$ and related Riemannian geometry  in local 
formulation. The
{\it isodual Riemannian spaces} are given by

\begin{eqnarray}
\Re^d(x^d,g^d,R^d) &:&\;\; x^d \;=\; \{-\hat{x}^\mu\} , \nonumber \\ &
&\;\; g^d\;=\; - g=(x) ,  \; g \in \Re(x,g,R)  , \nonumber \\ &  &\;\;
I^d \;=\; diag. ( -1, -1, -1, -1 )
% (2.35)
\label{eq:two-35}\end{eqnarray}
with interval  $x^{2d}$  $\;=\;$ $[x^{dt}$ $ \times^d  g^d(x^d)$
$\times^d
x^d]\times I^d$ = $[x^t$ $ \times  g^d(x^d)$ $\times
x]$ $\times I^d$ on $R^d$, where t stands for transposed.

        The {\it isodual Riemannian  geometry}   is the geometry    of
spaces $\Re^d$  over  $R^d$, and it is   also given by  step--by--step
isodualities  of  the     conventional    geometry,  including,   most
importantly, the isoduality of the differential and exterior calculus.

As an example, an {\it isodual vector field}  $X^d(x^d)$ on $\Re^d$ is
given  by   $X^d(x^d)\;=\;  -X(-x)$.   The  {\it    isodual   exterior
differential} of $X^d(x^d)$ is given by

\begin{equation}
D^dX^{kd}(x^d) \;=\; d^dX^{kd}(x^d) + \Gamma^d_i{}^k_j \times^d X^{id}
\times^d d^dx^d \;=\; D X^k(-x),
%(2.36)
\label{eq:two-36}\end{equation}
where the    $\Gamma^d$'s  are the  components  of    the {\it isodual
connection}. The {\it isodual covariant derivative} is then given by

\begin{equation}
X^{id}(x^d)_{|^  {dk}}  \;=\;  \partial^d  X^{id}(x^d)  /^d \partial^d
x^{kd} + \Gamma^d_i{}^k_j \times^d X^{jd}(x^d) \;=\; - X^i(-x)_{|^ k}.
% (2.37)
\label{eq:two-37}\end{equation}

The   interested reader can then easily   derive the isoduality of the
remaining notions of the conventional geometry.

        It is an   instructive exercise for  the interested  reader to
work out in detail the proof of the following:

\vskip 0.5cm {\bf Lemma 2.3:}  {\sl The  isoduality of the  Riemannian
space      $\Re(x,g,R)$    to      its    anti--automorphic     image
${\Re}^d(x^d,g^d,R^d)$  is characterized by    the
following isodual quantities:}

\begin{equation}
\begin{array}{ll}
Basic unit & I \;\; \rightarrow \;\; I^d \;=\; -  I,\\ Metric & g \;\;
\rightarrow  \;\;    g^d \;=\; -   g,   \\  Connection  coefficients &
\Gamma_{klh} \;\; \rightarrow \;\; \Gamma^d_{klh} \;=\; - \Gamma_{klh}
,\\ Curvature tensor & R_{lijk} \;\; \rightarrow \;\; R^d_{lijk} \;=\;
- R_{lijk}, \\ Ricci tensor & R_{\mu\nu} \;\; \rightarrow R^d_{\mu\nu}
\;=\;- R_{\mu\nu}, \\
% (2.38)
Ricci scalar & R \;\;\rightarrow \;\;R^d \;=\; R  , \\ Einstein tensor
&G_{\mu\nu} \;\;\rightarrow  \;\;G^d_{\mu\nu}  \;=\;  - G_{\mu\nu}, \\
Electromagnetic    potentials&   A_\mu\;\;\rightarrow\;\;A^d_\mu\;=\;-
A_\mu,            \\                    Electromagnetic          field
&F_{\mu\nu}\;\;\rightarrow\;\;F^d_{\mu\nu}\;=\;-  F_{\mu\nu},  \\  Elm
energy-mom.   tensor&T_{\mu\nu}\;\;\rightarrow\;\;T^d_{\mu\nu}\;=\;  -
T_{\mu\nu},
\end{array}
\label{eq:two-38}\end{equation}

        The reader should be aware that recent studies have identified
the  universal  symmetry of  conventional gravitation  with Riemannian
metric  $g(x)$,    the  so--called  {\it    isoPoincar\`{e}  symmetry}
$\hat{P}(3.1)  \;=\; \hat{L}(3.1)\times \hat{T}(3.1)$~[6].  The latter
symmetry  is the image of  the  conventional symmetry constructed with
respect to the generalized unit

\begin{equation}
        \hat{I}(x) \;=\;[T(x)]^{-1}, %(2.39)
\label{eq:two-39}\end{equation}
where $T(x)$  is    a   $4\times 4$   matrix  originating   from   the
factorization of the Riemannian metric into the Minkowskian one,

\begin{equation}
g(x) \;=\; T(x)\times \eta.  % (2.40)
\label{eq:two-40}\end{equation}

In particular, since $T(x)$  is always positive--definite, we have the
local isomorphism $\hat{P}(3.1) \approx P(3.1).$

        The same Ref.~[6] has constructed  the operator version of the
{\it isodual    isoPoincar\`{e}   symmetry}  $\hat{P}^d(3.1)   \approx
P^d(3.1)$, whose classical realization is   the universal symmetry  of
the isodual Riemannian spaces $\Re^d$ over $R^d$.

        In summary, the geometries significant in this paper are:
\begin{itemize}
\item[{\bf 1)}]  {\bf   The conventional Euclidean,    Min\-kow\-ski\-an
and
Rie\-man\-ni\-an ge\-o\-me\-tri\-es,} which are used for the
characterization of
matter; and\\
\item[{\bf 2)}] {\bf The isodual Euclidean, Minkowskian and Riemannian
geometries,} which are used  for the characterization of antimatter.\\
\end{itemize}

\section{CLASSICAL ISODUAL  THEORY  OF  ANTIMATTER}
\setcounter{equation}{0}
\subsection{Fundamental assumption.} As it is well known, the
contemporary treatment of matter is characterized by {\it conventional
mathematics}, here referred  to conventional numbers,  fields, spaces,
etc. with  {\it positive  unit  and  norm}, thus having   conventional
positive characteristics of mass, energy, time, etc.

        In this paper we study the following:

\vskip 0.5cm {\bf  Hypothesis 3.1:}  {\sl Antimatter is characterized by
the isodual  mathematics,  that with isodual numbers,  fields, spaces,
etc., thus  having      negative--definite  units  and   norm.     All
characteristics  of   matter  therefore  change  sign  for  antimatter
represented via isoduality.}
\vskip 0.5cm

        The   above   hypothesis    evidently   provides  the  correct
conjugation  of the charge at the  desired classical level. However, by
no means, the sole change of the  sign of the  charge is sufficient to
ensure a consistent classical representation of antimatter. To achieve
consistency, the  theory must resolve  the main problematic  aspect of
current classical treatments     of antimatter, the fact   that   their
operator  image is not   the correct  charge  conjugation  of that  of
matter, as   evident from  the   existence  of  a single  quantization
procedure.

It appears that the above problematic aspect is indeed resolved by the
isodual theory. The main reason is that,  jointly with the conjugation
of  the charge, isoduality  also conjugates  {\it  all} other physical
characteristics   of matter.  This  implies    {\it two} channels   of
quantization, the conventional one  for matter and  a new {\it isodual
quantization} for antimatter (see App. A) such that its operator image
is indeed the charge conjugate of that of matter.

         In  this section we  shall study the  physical consistency of
the  theory  in    its  classical  formulation.  The    novel  isodual
quantization, the equivalence of isoduality and charge conjugation and
related operator issues are studied in papers~[5,10].

        To begin our analysis, we note that Hypothesis 3.1 removes the
traditional obstacles against  negative energies and masses. In  fact,
{\it particles with negative masses and  energies referred to negative
units are  fully equivalent   to  particles with positive  energy  and
masses referred to positive units}. Moreover, as we shall see shortly,
particles with negative energy referred  to negative units behave in a
fully physical way. This  has permitted the  study in ref.~[10] of the
possible elimination of necessary  use of second quantization for the
quantum characterization of antiparticles, as the reader should expect
because our  main   objective  is  the  achievement  of  equivalent
treatments for  particles and antiparticles at  {\it all levels}, thus
including first quantization.

 Hypothesis 3.1 also resolves the  additional, well known, problematic
aspects of motion backward in time. In fact, {\it time moving backward
referred to a negative unit is fully equivalent to time moving forward
referred  to a positive unit}.  This  confirms the plausibility of the
first conception of antiparticles by Stueckelberg and others as moving
backward   in  time (see  the   historical  analysis of Ref.~[2]), and
creates new  possibilities for  the  ongoing research on  the so-called
''time machine'' to be studied in separate works.

In this  section  we   construct  the  classical isodual  theory    of
antimatter at the   Galilean, relativistic and  gravitational  levels,
prove  its axiomatic consistency  and   verify its compatibility  with
available classical  experimental  evidence  (that  on electromagnetic
interactions only).  We  also identify the  prediction of  the isodual
theory  that antimatter  in   the field of
matter experiences gravitational repulsion (antigravity),
and point out the ongoing efforts  for its future experimental
resolutions~[12,15].   For     completeness,   the  classical  isodual
Lagrangian and Hamiltonian mechanics  are provided in the Appendix  as
the foundation of the isoquantization of the joining paper~[10].

\subsection{Representation of antimatter via the classical isodual
Galilean relativity.}  We now   introduce  the {\it  isodual  Galilean
relativity}  as  the    most    effective way  for     the   classical
nonrelativistic characterization of antimatter according to Hypothesis
3.1.

The    study can be   initiated  with the    isodual representation of
antimatter at the most   primitive dynamical level, that   of Newton's
equation. Once a complete symmetry between the treatment of matter and
antimatter is  reached  at  the  Newtonian level,  it is   expected to
persist at all subsequent levels.

The conventional {\it Newton's equations} for a system of N point-like
{\it particles} with (non-null) masses $m_a,$ $a \;=\;  1, 2, ..., N,$
in exterior conditions in vacuum are given by the familiar expression

\begin{equation}
m_a \;\times\; d\; v_{ka}\: /\: dt \;=\; F_{ka}(t, r, v) ,\;\; r \;=\;
\{ x, y, z \},\;\; a \;=\; 1, 2, ..., N, \;\; v \;=\; dr / dt ,
%(3.1)
\label{eq:three-1}\end{equation}
defined on  the  7-dimensional Euclidean space   $E_{Tot}(t,r,v) \;=\;
E(t,R_t)\times    E(r,\delta,R_r)\times     E(v,\delta,R_v)$      with
corresponding  7-dimensional   total   unit $I_{Tot}  \;=\;  I_t\times
I_r\times I_v$, where one usually  assumes $R_r \;=\; R_v,\; I_t \;=\;
1,\; I_r \;=\; I_v \;=\; Diag.  (1, 1, 1).$

        The {\it  isodual  Newton equations}  here submitted  for  the
representation of n point-like  antiparticles in vacuum are defined on
the isodual space

\begin{equation}
E^d(t^d,r^d,v^d) \;=\;E^d(t^d,R_t^d)\times E^d(r^d,\delta^d,R^d)\times
E^d(v^d,\delta^d,R^d) ,
%(3.2)
\label{eq:three-2}\end{equation}
with total  isodual  unit $I_{Tot}^d \;=\;   I_t^d\times
I_r^d\times I_v^d,\;
I_t^d \;=\;  -1,\; I_r^d \;=\; I_v^d \;=\;=  - Diag.  (1,  1, 1)$, and
can be written   for (non-null) {\it isodual  masses}   $m^d_a \;=\; -
m_a)$

\begin{equation}
m^d_a  \times^d d^d v^d_{ka} /^d d^d  t^d \;=\; F^d_{ka}(t^d,r^d, v^d)
,\; k \;=\; x, y, z,\; a \;=\; 1, 2, ..., N.
%(3.3)
\label{eq:three-3}\end{equation}
It  is easy    to see that,    when  projected in   the original space
$S(t,r,v),$   isoduality  changes     the    sign  of  all    physical
characteristics, as expected.  It is also easy  to see  that the above
isodual  equations are anti-isomorphic to   the conventional forms, as
desired.

We now introduce the {\it isodual Galilean symmetry} $G^d$(3.1) as the
step-by-step isodual image of the conventional symmetry $G$(3.1) (see,
e.g.,  Ref.~[16]). By using  conventional   symbols for the   Galilean
symmetry of  a system of  N  particles with  non-null masses $m_a,\; a
\;=\; 1,  2,  ...,  N, G^d(3.1)$  is  characterized  by  {\it  isodual
parameters and generators}

\begin{equation}
w^d \;=\; (\theta_k^d,  r_o^{kd}, v_{o}^{kd},  t_o^d)  \;=\; -w  ,\;\;
J_k^d \;=\;  \sum {}_a{}_{ijk} r_{ja}^d \times^d  p_{ja}^k \;=\; - J_k
,\;\; P_k^d \;=\; \sum {}_a p_{ka}^d \;=\; - P_k ,
%(3.4a)
\label{eq:three-4a}\end{equation}
\begin{equation}
G_k^d \;=\; \sum {}_a ( m_a^d  \times^d r_{ak}^d - t^d \times p_{ak}^d
),\;\; H^d  \;=\; \frac{1}{2}^d  \times^d \sum {}_a  p_{ak}^d \times^d
p_a^{kd} + V^d(r^d) \;=\; - H ,
% (3.4b)
\label{eq:three-4b}\end{equation}
equipped with the {\it isodual commutator} $(A.11)$, i.e.,

\begin{eqnarray}
\lefteqn{ [ A^d ,  B^d ]^d \;=\; \sum  {}_{a, k} [ (\partial^d A^d /^d
\partial^d r_a^{kd} ) \times^d (\partial^d B^d /^d \partial^d p_{ak}^d
) -} \nonumber  \\ & &  - (\partial^d B^d  /^d \partial^d r_{a}^{kd} )
\times^d (\partial^d A^d /^d \partial^d p_{ak}^d )] \;=\;  - [ A , B ]
.
%(3.5)
\label{eq:three-5}\end{eqnarray}

In accordance with rule~(\ref{eq:two-26}), the structure constants and
Casimir invariants of    the  isodual  Lie algebra  $G^d(3.1)$     are
negative--definite.  From  rule~(\ref{eq:two-27}),   if $g(w)$  is  an
element  of the (connected component)  of  the Galilei group $G(3.1)$,
its isodual is characterize by

\begin{equation}
g^d(w^d) \;=\; e^{d^{-i^d\times^d w^d\times^d  X^d}} \;=\; - e^{i\times
(-w)\times X} \;=\; - g(-w) \in G^d(3.1).
%(3.6)
\label{eq:three-6}\end{equation}

The {\it isodual Galilean transformations} are then given by

\begin{equation}
t^d  \;\;\rightarrow\;\; t '{}^d  \;=\; t^d +  t_o^d \;=\; - t ', \;\;
r^d \;\;\rightarrow\;\; r '{}^d \;=\; r^d + r_o^d \;=\; - r '
%(3.7a)
\label{eq:three-7a}\end{equation}
\begin{equation}
r^d\;\;  \rightarrow  \;\; r  '{}^d \;=\; r^d   + v_o^d \times^d t_o^d
\;=\; - r  ',\;\; r^d \;\;\rightarrow\;\;  r '{}^d \;=\; R^d(\theta^d)
\times^d r^d \;=\; - R(-\theta) .
%(3.7b)
\label{eq:three-7b}\end{equation}
where  $R^d(\theta^d)$ is an  element   of the {\it  isodual  rotational
symmetry} first studied in the original proposal~[3].

The desired classical  nonrelativistic characterization of  antimatter
is therefore  given by imposing  the $G^d(3.1)$ invariance  of isodual
equations~(\ref{eq:three-3}). This  implies, in  particular, that  the
equations   admit a  representation   via the  isodual Lagrangian  and
Hamiltonian mechanics outlined in Appendix A.

We now  verify that the above  isodual representation of antimatter is
indeed consistent with available  classical experimental knowledge for
antimatter,    that  under electromagnetic    interactions.  Once this
property is    established  at the    primitive  Newtonian level,  its
verification at  all subsequent levels of  study is expected from mere
compatibility arguments.

        Consider a conventional, classical, massive {\it particle} and
its {\it antiparticle} in exterior  conditions in vacuum. Suppose that
the particle and antiparticle  have charge $-e$ and $+e$, respectively
(say, an  {\it electron} and  a {\it  positron}), and  that they enter
into the gap of a magnet with constant magnetic field {\bf B}.

As it is well known,  visual experimental observation establishes that
particles and antiparticles  have spiral trajectories of {\it opposite
orientation}. But this behavior occurs  for the {\it representation of
both   the  particle and   its   antiparticle  in  the same  Euclidean
space}. The situation under  isoduality is different, as described  by
the following:

\vskip  0.5cm  {\bf  Lemma 3.1:}  {\sl   The trajectory  of  a charged
particle in Euclidean space under  a magnetic field and the trajectory
of the corresponding  antiparticle in isodual Euclidean space coincide.}
\vskip 0.5cm

{\bf Proof:}  Suppose  that the  particle  has  negative charge   $-e$
in
Euclidean  space $E(r,\delta,R)$, that is, the  value  $-e$ is defined
with respect to the positive unit $+1$ of the underlying field of real
numbers $R  \;=\; R(n,+,\times )$.  Suppose that the particle is under
the influence of the  magnetic field {\bf  B}. The characterization of
the corresponding antiparticle via  isoduality implies the reversal of
the  sign of all physical quantities,  thus yielding the charge $(-e)^d
\;=\; +e$  in the isodual  Euclidean space $E^d(r^d,\delta^d,R^d)$, as
well as the reversal  of the magnetic  field $B^d \;=\;  -B$, although
now defined with respect to  the negative unit  $(+1)^d \;=\; -1$.  It
is then evident that the trajectory of  a particle with charge $-e$ in
the field B  defined with respect to  the unit $+1$ in Euclidean space
and that for the antiparticle of charge $+e$ in the field $-B$ defined
with   respect  to  the   unit     -1  in  isodual   Euclidean   space
coincide. q.e.d.

An aspect  of Theorem  3.1 which  is  particularly important for  this
paper is given by the following

\vskip 0.5cm  {\bf Corollary 3.1.A:}  {\sl Antiparticles reverse their
trajectories  when  projected  from   their isodual  space  into   the
conventional space.}

 Lemma 3.1 assures that isodualities permit  the representation of the
correct trajectories  of antiparticles as physically observed, despite
their negative energy, thus providing the foundations for a consistent
representation   of  antiparticles   at   the  level of   {\it  first}
quantization studied in paper~[10]. Moreover,  Lemma 3.1 tells us that
the  trajectories of antiparticles  may {\it  appear} to  exist in our
space  while in reality they may  belong to  an independent space, the
isodual Euclidean space, coexisting with our own space.

        To verify the validity  of the isodual theory  at the level of
Newtonian laws  of electromagnetic phenomenology,  let us consider the
{\it  repulsive} Coulomb force among  two  {\it particles of negative}
charges $-q_1$ and $-q_2$ in $E(r,\delta,R)$,

\begin{equation}
F \;=\; K \times (-q_1) \times (-q_2)/ r\times r > 0 ,
%(3.8)
\label{eq:three-8}\end{equation}
where the operations of multiplication $\times  $ and division $/$ are
the conventional ones of the underlying field $R(n,+,\times )$.  Under
isoduality to $E^d(r^d,\delta^d,R^d)$ we have

\begin{equation}
F^d \;=\; K^d \times^d (-q_1)^d \times^d (-q_2)^d /^d r^d \times^d r^d
\;=\; - F < 0 ,
%(3.9)
\label{eq:three-9}\end{equation}
where $\times^d \;=\; -\times   $ and $/^d  \;=\;  -/$  are the
isodual operations of the underlying field $R^d(n^d,+,\times^d)$.

But the isodual force  $F^d \;=\; -F$ occurs  in the isodual Euclidean
space and it  is therefore defined  with respect to  the unit -1. As a
result, isoduality correctly represents  the {\it repulsive} character
of the Coulomb force  for two {\it  antiparticles} with {\it positive}
charges.

The Coulomb force  between a {\it particle}  and an {\it antiparticle}
can  only  be computed  by  {\it projecting   the antiparticle in  the
conventional space of the particle or  vice-versa}. In the former case
we have

\begin{equation}
F \;=\; K \times (-q_1) \times (-q_2)^d / r\times r < 0 ,
%(3.10)
\label{eq:three-10}\end{equation}
thus  yielding   an    {\it  attractive}   force,  as   experimentally
established. In the projection of the particle in the isodual space of
the antiparticle we have

\begin{equation}
F^d \;=\; K^d \times^d (-q_1) \times^d (-q_2)^d /^d r^d \times^d r^d >
0 .
% (3.11)
\label{eq:three-11}\end{equation}

But this force is  now referred to  the unit -1,  thus resulting to be
again {\it attractive}.

In conclusion, the isodual Galilean relativity correctly represent the
electromagnetic interactions of  antimatter at the classical Newtonian
level.

\subsection{Representation of antimatter via the isodual special
relativity. } We now introduce the {\it isodual special relativity} as
the best way to represent  classical relativistic antimatter according
to Hypothesis 3.1.

In essence, the conventional  special  relativity (see, e.g.,  Pauli's
historical    account~[17])  is    constructed     on the  fundamental
4~--~di\-men\-sio\-nal unit of the Minkowski space $I \;=\;$ $ Diag. $
$\{1, 1, 1\}, 1)$,  which represents the  dimensionless units of space
$\{+1, +1, +1\}$, and  the dimensionless unit of  time +1, and is  the
unit of  the   Poincar\`{e}  symmetry  P(3.1).   The  isodual  special
relativity is characterized by the map

\begin{equation}
I \;=\; diag. (\{1, 1, 1\}, 1 ) > 0 \;\;\; \rightarrow \;\;\; I^d\;=\;
   - diag. (\{1, 1, 1\}, 1 ) < 0 .
%(3.12)
\label{eq:three-12}\end{equation}
namely,  it is based on  {\it negative units  of  space and time}. The
isodual special relativity is then  expressed by the isodual image  of
{\it  all} mathematical  and  physical   aspects of  the  conventional
relativity in  such a way to admit  the negative--definite quantity
$I^d$
ass the correct left and right unit.

       This implies  the reconstruction  of the entire  mathematics of
the  special   relativity  with   respect  to   the    single, common,
4-dimensional  unit $I^d$, including:   the {\it  isodual field}  $R^d
\;=\; R^d(n^d,+^d,\times^d)$  of   {\it isodual   numbers} $n^d  \;=\;
n\times I^d  \;=\;  - n\times  I$   with fundamental unit $I^d   \;=\;
-Diag.(1,  1,   1=  ,    1);$  the    {\it isodual   Minkowski  space}
$M^d(x^d,\eta^d,R^d)$ with isodual coordinates $x^d \;=\; x\times I^d$,
isodual
metric $\eta^d \;=\; -\eta$ and basic invariant over $R^d$

\begin{equation}
( x -  y  )^{d2 d} \;=\; [(x^\mu -  y^\mu ) \times  \eta_{\mu\nu}^d
\times ( x^\nu - y^\nu ) \times I^d \in R^d;
%(3.13)
\label{eq:three-13}\end{equation}
the fundamental
{\it isodual Poincar\`{e}} symmetry~[6]

\begin{equation}
P^d(3.1) \;=\; L^d(3.1) \times^d T^d(3.1) ,
%(3.14)
\label{eq:three-14}\end{equation}
where $L^d(3.1)$ is the  {\it isodual Lorentz symmetry}, $\times^d$ is
the {\it isodual  direct  product} and $T^d(3.1)$ represents  the {\it
isodual  translations}, whose  classical   formulation is given  by  a
simple relativistic extension of the  isodual Galilean symmetry of the
preceding section.

The algebra   of the connected   component $P_{+}^{\uparrow   d}(3.1)$
of
$P^d(3.1)$ can be constructed in  terms of the isodual parameters $w^d
\;=\; \{-w_k\} \;=\; \{-\theta, -v,  -a\}$ and isodual generators $X^d
\;=\; -X  \;=\;  \{-X_k\} \;=\;  \{-M_{\mu\nu},  -P_\mu\},$ where  the
factorization by  the  four-dimensional  unit  I  is  understood.  The
isodual commutator rules are given by

\begin{equation}
[ M_{\mu\nu}^d,     M_{\alpha\beta}^d   ]^d     \;=\;i^d      \times^d
(\eta_{\nu\alpha}^d   \times^d    M_{\mu\beta}^d  - \eta_{\mu\alpha}^d
\times^d M_{\nu\beta}^d - \eta_{\nu\beta}^d \times^d M_{\mu\alpha}^d +
\eta_{\mu\beta}^d \times^d \hat{M}_{\alpha\nu}^d ),
%(3.15a)
\label{eq:three-15a}\end{equation}
\begin{equation}
[    M_{\mu\nu}^d   ,      p_{\alpha}^d  ]^d    \;=\;  i^d    \times^d
(\eta_{\mu\alpha}^d  \times^d  p_\nu^d  - \eta_{\nu\alpha}^d  \times^d
p_\mu^d ), \;\;\; [p_\alpha^d , p_\beta^d ]^d \;=\; 0 ,
%(3.15b)
\label{eq:three-15b}\end{equation}
The isodual group $P_+^{\uparrow  d}(3.1)$ has a structure
  similar to   that  of Eq.s~(\ref{eq:three-6}).   These  results then
  yield the following

\vskip 0.5cm {\bf Lemma 3.2:}  {\sl The classical isodual Poincar\`{e}
transforms are given by}

\begin{eqnarray}
x^{1d}{}' \;&=&\; x^{1d} = - x^1 ,\;\; x^{2d}{}' \;=\; x^{2d} = -x^2  ,
\nonumber \\ x^{3d}{}' \;&=&\;
\gamma^d \times^d ( x^{3d} -  \beta^d \times^d x^{4d} ) = - x^3{}',
\;\; x^{4d}{}'  \;=\; \gamma^d \times^d ( x^{4d}  -
\beta^d \times^d x^{3d} ) = -x^4{}' ,\nonumber \\
x^{d\mu \prime}\;&=&\; x^{d\mu} + a^{d\mu\: d} \;=\; -x^{\mu}{}',
\nonumber  \\ x^{d\mu \prime}\;&=&\; \pi^d \times^d
x^d \;=\; -\pi \times  x \;=\; \-( -r,  x^4 ),\;\; \tau^d \times^d x^d
\;=\;
-\tau \times x \;=\; -( r , - x^4 ) ,
%(3.16)
\label{eq:three-16}\end{eqnarray} {\it where}

\begin{equation}
\beta^d  \;=\; v^d /^d  c_o^d \;=\; -\beta  ,\;\;  \beta^{d2d} \;=\; -
\beta^2, \;\;\gamma^d \;=\; - (1 - \beta^2 )^{-1/2} .
%    (3.17)
\label{eq:three-17}\end{equation}
{\it and the  use of the isodual  operations  (quotient, square roots,
etc.), is implied}.
\vskip 0.5cm

        The {\it isodual  spinorial  covering} of the  Poincar\`{e}
symmetry $\cal{P}^d$(3.1) = $SL^d(2.C^d)\times^d T^d(3.1)$ can then
be constructed via the same methods.

The basic  postulates  of the  isodual special  relativity are also  a
simple isodual image of the conventional postulates. For instance, the
{\it  maximal isodual causal  speed} is the   speed of light in $M^d$,
i.e.,

\begin{equation}
V_{max} \;=\; c_o^d \;=\; - c_o ,
%(3.18)
\label{eq:three-18}\end{equation}
with the     understanding     that it is      referred    to  a  {\it
negative--definite  unit},   thus  being    fully equivalent   to  the
conventional  maximal speed $c_o$ referred to  a positive unit. A
similar
situation occurs for all other postulates.

        A fundamental property of the isodual theory is the following:

\vskip 0.5cm  {\bf  Theorem  3.1:} {\it  The line  elements of    metric
or
pseudo-metric   spaces are  isoselfdual  (Definition 2.1),  i.e., they
coincide with their  isodual images. In  particular, isoduality leaves
invariant  the fundamental   space-time   interval   of   the  special
relativity,}

\[
x^{d\:2\:d} \;=\;  (  x^{\mu d}\times^d  \eta_{\mu\nu}^d\times^d  x^{\nu
d})
\;=\;
\]
 \begin{equation} \;=\; (  - x^1 x^1 -  x^2 x^2 - x^3  x^3 - x^4 x^4 )
\times ( - I ) \equiv ( x^1 x^1 + x^2 x^2 +  x^3 x^3 - x^4 x^4 )\times
I \;=\; x^2 .
% (3.19)
\label{eq:three-19}\end{equation}

        The above  novel property evidently assures  that conventional
relativistic laws for matter are also valid for antimatter represented
via  isoduality,  since they  share   the same fundamental  space-time
interval.

       The  above property   illustrates  that the isodual   map is so
natural to creep in un-noticed. The reason  why, after about a century
of studies,  the    isoduals of the  Galilean,  special   and  general
relativities escaped  detection is that  their identification required
the prior knowledge of {\it new numbers}, those with a negative unit.

        Note that the use of the {\it  two} Minkowskian metrics $\eta$
and $\eta^d   \;=\;=  -\eta$ has     been popular since    Minkowski's
times. The point  is that both metrics are  referred to the {\it same}
unit I, while in the isodual theory one metric is referred to the unit
I on the field $R(n,+,\times )$ of conventional numbers, and the other
metric is referred  to the new unit $I^d  \;=\;  -I$ on the new  field
$R^d(n^d,+^d,\times^d)$ of isodual numbers $n^d \;=\; n\times I^d$.

The novelty of  the     isodual relativities is illustrated    by  the
following

        \vskip 0.5cm {\bf  Lemma  3.3.}  {\it Isodual maps  and
space-time
inversions are inequivalent.}
\vskip 0.5cm

        In fact, space-time inversions are characterized by the change
of  sign  $x \rightarrow  -x$  {\it by always  preserving the original
metric  referred to positive units},  while isoduality implies the map
$x  \rightarrow x^d  \;=\;  -x$ {\it but   now referred to an  isodual
metric} $\eta^d  \;=\; -\eta$ {\it   with negative units}  $I^d  \;=\;
-I$.  Thus, {\it space-time inversions occur   in the same space while
isoduality implies the map to a different space}. Moreover,
as shown by lemma 3.2 isodualities interchange the space and time
inversions.

        The  interested   reader is   encouraged to  verify  that  the
physical   consistency   in   the   representation  of electromagnetic
interactions by the  isodual Galilean relativity  carries  over in its
entirety at  the  level   of  the  isodual  special relativity,   thus
confirming the  plausibility of the  isodual theory of antimatter also
at the classical relativistic level.

\subsection{Representation of antimatter via the isodual general
relativity. }   We  finally   introduce  the   {\it   isodual  general
relativity} as the most   effective gravitational characterization  of
antimatter  according to  Hypothesis   3.1.  The  new  image is   also
characterized   by the isodual   map   of {\it all}   aspects of   the
conventional relativity (see,  e.g.,~[18]), now defined  on  the
isodual  Riemannian
spaces $\Re^d(x^d,g^d,R^d)$ of Sect. 2.7

        The  primary motivation warranting the  study of the above new
image  of  general    relativity  is  the  following.  A
problematic   aspect in the  use  of the  Riemannian  geometry for the
representation of   {\it antimatter}  is  the {\it  positive--definite
 energy-momentum tensor}.

        In fact,  such a gravitational representation  of
antimatter   has  an
operator image which  is not the  charge conjugate of that  of matter,
does not admit the  negative-energy solutions  as needed for  operator
treatments of antiparticles, and may be one of the reasons for the lack
of achievement until now  of a consistent grand unification  inclusive
of gravitation. Afetr all,  gauge theories are bona-fide  field theories
which,  as such, admit  both positive-  and negative-energy solutions,
while   the  contemporary    formulation    of gravity  admits    only
positive-energy states, with an evident structural incompatibility.

        Isoduality offers a new possibility for a future resolution of
these  shortcomings.  In  fact, the  isodual  Riemannian  geometry  is
defined  on the isodual field  of real numbers $R^d(n^d,+^d,\times^d)$
for      which      {\it   the     norm      is   negative--definite},
Eq.~(\ref{eq:two-11}).   As a  result, {\it  all  quantities which are
positive in Riemannian geometry become negative under isoduality, thus
including the energy-momentum tensor}.

        Explicitly, the electromagnetic  field follows the simple rule
under isoduality

        \begin{equation}  F^d_{\mu\nu}   \;=\;   \partial^dA^d_\mu /^d
\partial^dx^{\nu d} - \partial^d A^d_\nu /^d \partial^d x^{d\mu} \;=\;
- F_{\mu\nu} ,
%(3.20)
\label{eq:three-20}\end{equation}
and for the energy-momentum tensor we have the corresponding law

        \begin{equation}   T^d_{\mu\nu}  \;=\; (4m)^{-1d}  \times^d  (
F^d_{\mu\alpha}  \times^d  F^d_{\alpha_\nu} + (1/4)^{-1d} \times^d
g^d_{\mu\nu}
\times^d   F^d_{\alpha\beta}   \times^d  F^{d\alpha\beta} )   \;=\;  -
T_{\mu\nu}.
%(3.21)
\label{eq:three-21}\end{equation}

        As such, antimatter represented in isodual Riemannian geometry
has {\it negative--definite energy-momentum tensor} and other physical
quantities, as desired, thus offering new
possibilities for attempting a grand unified theory.

        For completeness, we mention  here the {\it  isodual Einstein}
equations for the {\it exterior gravitational problem of antimatter in
vacuum}

        \begin{equation}  G_{\mu\nu}^d    \;=\;       R_{\mu\nu}^d   -
\frac{1}{2}^d \times^d g^d_{\mu\nu} \times^d R^d \;=\; k^d \times^d
T^d_{\mu\nu} ,
%(3.22)
\label{eq:three-22}\end{equation}

        We also mention the field equations  characterized by the {\it
Freud  identity}~[19]    of the    Riemannian  geometry (reviewed   by
Pauli~[17] and then generally forgotten)

\begin{equation}
R^\alpha_\beta -  \frac{1}{2}  \times \delta^\alpha_\beta  \times R  -
\frac{1}{2} \times      \delta^\alpha_\beta   \times  \Theta     \;=\;
U^\alpha_\beta + \partial  V^{\alpha\rho}_\beta/\partial x^\rho \;=\;  k
\times  (
t^\alpha_\beta - \tau^\alpha_\beta )
%(3.23)
\label{eq:three-23}\end{equation}
  where

\begin{equation}
\Theta     \;=\;     g^{\alpha\beta}       g^{\gamma\delta}          (
\Gamma_{\rho\alpha\beta}          \Gamma_\gamma^\rho{}_\beta         -
\Gamma_{\rho\alpha\beta} \Gamma_\gamma^\rho{}_{\delta}) ,
%(2.24a)
\label{eq:two-24a}\end{equation}

\begin{equation}
U^\alpha_\beta   \;=\; - \frac{1}{2}\frac{\partial    \Theta}{\partial
g^{\alpha\beta}_{|_\alpha}}\hat{g}^{\alpha\beta}\uparrow_\beta ,
%(2.24b)
\label{eq:two-24b}\end{equation}

\[
V^{\alpha\rho}_\beta\;=\;\frac{1}{2}[g^{\gamma\delta}(\delta^\alpha_\beta
\Gamma_\alpha^\rho{}_\equiv         -                \delta^\rho_\beta
\Gamma_\gamma^\rho{}_\delta ) +
\]
\begin{equation}
+  (   \delta^\rho_\beta  g^{\alpha\gamma}    -    \delta^\alpha_\beta
g^{\rho\gamma}  )    \Gamma_\gamma^\delta{}_\delta    + g^{\rho\gamma}
\Gamma_\beta^\alpha{}_\gamma              -           g^{\alpha\gamma}
\Gamma_\delta^\rho{}_\gamma] ,
% (3.24c)
\label{eq:three-24c}\end{equation}

which  are currently under study   for the {\it interior gravitational
problem of matter}.

        The isodual version of Eq.s~(\ref{eq:three-23})

\begin{equation}
R^\alpha_\beta{}^d    -  \frac{1}{2}^d \times^d \delta^\alpha_\beta{}^d
\times^d R^d - \frac{1}{2}^d \times^d \delta^\alpha_\beta{}^d \times^d
\Theta^d \;=\; k^d \times^d ( t^\alpha_\beta{}^d -
\tau^\alpha_\beta{}^d)
%(3.25)
\label{eq:three-25}\end{equation}
are then  suggested  for  the  study  of  {\it interior  gravitational
problems of antimatter}.

        It is instructive for the interested reader to verify that the
preceding physical  consistency of the isodual  theory carries over at
the above   gravitational   level,  including the   {\it   attractive}
character of antimatter-antimatter systems  and their correct behavior
under electromagnetic interactions.

        Note  in  the latter respect that    {\it curvature in isodual
Riemannian spaces  is  negative--definite} (Sect. 2.7).  Nevertheless,
such negative value for antimatter-antimatter systems is referred to a
negative unit, thus resulting in attraction.

        The universal symmetry of the  isodual general relativity, the
{\it    isodual isoPoincar\`{e}    symmetry} $\hat{P}^d(3.1)   \approx
P^d(3.1)$, has been introduced at the  operator level in Ref.~[6]. The
construction of its classical counterpart is straightforward, although
it cannot   be  reviewed here  because  it requires  the  broader {\it
isotopic   mathematics},       that       based  on        generalized
unit~(\ref{eq:two-39}).

\subsection{The prediction of antigravity.} We close  this   paper with
the
indication that the  {\it  isodual theory  of antimatter predicts  the
existence of antigravity (here defined as  the reversal of the sign of
the curvature tensor in our  space--time) for antimatter in the  field
of matter, or vice--versa}.

        The  prediction originates at  the  primitive Newtonian level,
persists at  all  subsequent levels of   study~[10],  and it  is  here
identified as a consequence  of the theory  without  any claim  on its
possible validity  due to the lack  of experimental knowledge  at this
writing on the gravitational behavior  of antiparticles.

        In essence, antigravity is  predicted by the interplay between
conventional geometries  and  their isoduals  and,  in particular,  by
Corollary 3.1.A according to  which  the trajectories we  observe  for
antiparticles are  the  {\it  projection}  in our  space--time  of the
actual  trajectories in isodual space.  The  use of the same principle
for the case of the gravitational field then yields antigravity.

        Consider the Newtonian gravitational force of two conventional
(thus positive) masses $m_1$ and $m_2$

\begin{equation}
F \;=\; - G \times m_1 \times m_2 / r \times r < 0 ,
%(3.26)
\label{eq:three-26}\end{equation}
where  the  minus sign    has     been added  for   similarity    with
law~(\ref{eq:three-8}).

        Within the context  of contemporary theories, the masses $m_1$
and   $m_2$ remain  positive irrespective  of   whether referred to  a
particle or an antiparticle. This yields the well known {\it Newtonian
gravitational attraction}  among   any pair  of  masses,   whether for
particle--particle,         antiparticle--antiparticle              or
particle--antiparticle.

        Under  isoduality  the situation   is  different.  First,  the
particle--particle   gravitational   force yields    exactly  the same
law~(\ref{eq:three-5}).  The  case of antiparticle--antiparticle under
isoduality yields the different law

\begin{equation}
F^d \;=\; - G^d \times^d m_1^d \times^d m_2^d /^d r^d \times^d r^d > 0
.
%(3.27)
\label{eq:three-27}\end{equation}

But  this force is defined with  respect to the  negative unit -1. The
isoduality  therefore correctly   represents   the  {\it   attractive}
character   of   the  {\it    gravitational}   force   among   two {\it
antiparticles}.

        The  case of  particle--antiparticle under isoduality requires
the {\it projection} of the antiparticle in the space of the particle,
as  it is the case for   the electromagnetic interactions of Corollary
2.1.A

\begin{equation}
F \;=\; - G \times m_1 \times m_2^d / r \times r > 0 ,
% (3.28)
\label{eq:three-28}\end{equation}
which is now {\it repulsive}, thus illustrating the prediction of
antigravity. Similarly, if we project  the particle in the space of
the antiparticle we have

\begin{equation}
F^d \;=\; - G^d \times^d m_1 \times^d m_2^d /^d r^d \times^d r^d < 0 ,
%(3.29)
\label{eq:three-29}\end{equation}
which is also repulsive because referred to the unit -1.

        We can summarize the   above results by  saying that  {\it the
classical  representation  of  antiparticles via  isoduality   renders
gravitational interactions  equivalent to the electromagnetic ones, in
the sense that the Newtonian  gravitational law becomes equivalent  to
the  Coulomb law.  Note the impossibility   of achieving these results
without isoduality}.

        The interested reader can verify  the persistence of the above
results at the relativistic and gravitational levels.

        We   should   indicate   that  the   electroweak   behavior of
antiparticles is  experimentally established  nowadays, while there no
final   experimental   knowledge  on   the gravitational   behavior of
antiparticles is available at this writing.

        A first experiment on the gravity of antiparticles was done by
Fairbank and  Witteborn~[20]  via low   energy  positrons in  vertical
motion,   although the measurements    were not conclusive because  of
interferences from stray fields and other reasons.

        Additional data on the gravity of antiparticles are those from
the LEAR machine on antiprotons at CERN~[21],  although these data too
are inconclusive because of the excessive energy of the antiprotons as
compared  to the    low   magnitude  of gravitational    effects,  the
sensitivity of the measures and other factors.

        Other  information on the  gravity    of antiparticles is   of
theoretical character, as reviewed, e.g., in Ref.~[22], which outlines
various arguments  {\it    against}  antigravity, such  as   those  by
Morrison,  Schiff and  Good  and others. The   latter arguments do not
apply under isodualities    owing to  their essential  dependence   on
positive units, as one can verify.

        The argument against antigravity based on the positronium~[22]
also do  not apply  under  isoduality, because systems  of  elementary
particle--antiparticle are attracted  in  both  fields of matter   and
antimatter under    the isodual  theory,   as  studies  in  the  joint
paper~[10].

        Review~[22] also  indicates   models in which the  gravity  of
antimatter in the  field of matter  is  {\it weakened, yet it  remains
attractive}.

        We  can therefore  state  that  the gravitational behavior  of
antiparticles is  theoretically  and experimentally unsettled  at this
writing.

        The   true  scientific  resolution    is  evidently that   via
experiments,  such as that proposed by   Santilli~[7] via the  use of a
suitably   collimated  beam  of very    low energy positrons  in  {\it
horizontal} flight in a vacuum  tube of sufficient length and diameter
to yield a resolutory answer, that is, a displacement under gravity at
the end of the flight up
or down  which is visible by the naked
eye.

        According to  Mills~[12],  the above experiment  appears to be
feasible with current technology via the use  of $\mu eV$ positrons in
a horizontal vacuum tube of about 100 m in  length and 1 m in diameter
for  which stray fields and patch  effects should be small as compared
to the gravitational  deflection. A number of  additional experimental
proposals to measure  the gravity  of  antiparticles are  available in
Proceedings~[15], although  their measures are  more sophisticated and
not "visible by the naked eyes" as test~[7,12].

        In   closing  we indicate  that  the   most forceful  argument
favoring the existence of antigravity is given by studies on the {\it
origin} of the gravitational field. In fact, the mass of all particles
constituting a    body  has a  primary  electromagnetic   origin, with
second--order contributions   from weak   and strong  interactions. By
using this established  physical evidence, Ref.~[23] proposed the {\it
identification}  (rather  than  the ''   unification '') {\it   of the
gravitational field with  the fields  originating  mass}. This can  be
done  by  identifying the  $\tau$--tensor in  Eqs.~(\ref{eq:three-23})
with  the   electromagnetic   field    originating   mass, and     the
$\tau$--tensor with the weak   and  strong contributions.  Since   the
latter are  short range,  in the  exterior problem  in vacuum we would
only  have   the   {\it   identification of  the    gravitational  and
electromagnetic fields} which would evidently imply the equivalence of
the respective     phenomenologies, thus including the    capability for
attractive and repulsive forces for both fields.

        Note that this implies the existence of a first--order nowhere
null  electromagnetic source  also for bodies  with  null total charge
(see Ref.~[23]  for details). Note also  the full compatibility of the
above argument with the isodual  representation of antimatter, because
both approaches   imply the equivalence of  Coulomb's  law for charges
with Newton's law for masses.

        The forceful nature  of the above  argument is due to the fact
that the  lack of antigravity  would  imply  the lack of  identity  of
gravitational  and electromagnetic interactions.  In turn,  this would
require a revision of  the contemporary theory of elementary particles
in such a  way to  {\it avoid} the  primary  electromagnetic origin of
their mass.

\vspace{2cm}   \appendix{\bf   APPENDIX  A.}   \setcounter{section}{1}
\def\thesection                                       {\Alph{section}}
\def\theequation{\thesection.\arabic{equation}}

%\large \bf         APPENDIX A.}
\vskip 0.5cm
         {\bf A.1: Isodual Lagrangian mechanics}.

After having     verified the isodual theory     of  antimatter at the
primitive  Newtonian level, it may  be of  some  value to outline its
analytic representation because it constitutes  the foundations of the
novel quantization for antimatter studied in the joint paper~[10].

        A  conventional   (first--order)  Lagrangian   $L(t,x,v) \;=\;
\frac{1}{2}mv^kv_k +  V(t,x,v)$  on the configuration  space $E(t,x,v)
\;=\;  E(t, R_t)\times E(r,\delta,R_r)\times  E(v,\delta,R_v)  $ of
Newton's
equations  is     mapped under isoduality   into   the  negative value
$L^d(t^d,r^d,v^d)    \;=\;  -   L$      defined   on   isodual   space
$E^d(t^d,r^d,v^d)$  of   Eq.~(\ref{eq:three-2}).   The   {\it  isodual
Lagrange equations} are then given by

\begin{equation}
\frac{d^d}{d^d   t^d}d\frac{\partial^d L^d(t^d, r^d,  v^d)}{\partial^d
v^{kd}}d - \frac{\partial^dL^d(t^d,  r^d, v^d)}{\partial^d r^{kd}} d =
0,
%(A.1)
\label{eq:a-1}\end{equation}
All various aspects of the {\it isodual Lagrangian mechanics} can then
be readily derived.

        It is easy to see  that Lagrange's equations change sign under
isoduality  and can  therefore provide  a {\it direct  representation}
(i.e.,  a representation without   integrating factors) of the isodual
Newton's equations,

\[
\frac{d^d}{d^dt^d}d\frac{\partial^d    L^d(t^d,  r^d, v^d)}{\partial^d
v^{kd}}d - \frac{\partial^d L^d(t^d, r^d, v^d)}{\partial^d x^{kd}}d =
\]
\begin{equation}
\;=\; m^d_k \times^d d^d  v^d_k /^d d^d t^d   - F^d_k{}^{SA}(t, r,  v)
\;=\; 0 .
%(A.2)
\label{eq:a-2}\end{equation}

Where SA stands for {\it variational selfadjointness},
i.e., verification of the conditions to be derivable from a potential.
The compatibility of the isodual Lagrangian mechanics with the
primitive Newtonian results then follows.

\vskip 0.5cm
{\bf A.2: Isodual Hamiltonian mechanics}.

                The {\it isodual Hamiltonian} is evidently given by

\begin{equation}
H^d \;=\; p^d_k \times^d p^{dk} /^d (2m)^d +  V^d(t^d, r^d, v^d) \;=\;
- H .
%(A.4)
\label{eq:a-4}\end{equation}

It can   be  derived from  (nondegenerate) isodual  Lagrangians  via a
simple isoduality of the Legendre transforms and  it is defined on the
7--dimensional carrier space (for one particle)

\begin{equation}
E^d(t^d,r^d,p^d)   \;=\;  E^d(t^d,R^d_t) \times  E^d(r^d,\delta^d,R^d)
\times E^d(p^d,\delta^d,R^d) .
%        (A.5)
\label{eq:a-5}\end{equation}

        The {\it isodual canonical action} is given by

\[
A^{\circ d}  \;=\; \int_{t_1}^{t_2} ( p^d_k \times^d
  d^d r^{kd} - H^d \times^d d^dt^d ) \;=\;
\]
\begin{equation}
\;=\; \int_{t_1}^{t_2} [ R^{\circ d}_\mu(b^d) \times^d
d^db^{\mu d} - H^d \times^d d^dt^d], \; R^{\circ} \;=\;  \{ p, 0 \},\;
b \;=\; \{ x, p \} ,\; \mu \;=\; 1, 2, ...  , 6.
%(A.6)
\label{eq:a-6}\end{equation}
Conventional variational techniques under simple isoduality then yield
the {\it isodual Hamilton equations} which  can be written in disjoint
form

\begin{equation}
\frac{d^d   x^{kd}}{d^d        t^d}    =  \frac{\partial^dH^d(t^d,x^d,
p^d)}{\partial^dp^d_k},\;\frac{d^d     p^d_k}{d^dt^d}=-\frac{\partial^d
H^d(t^d,x^d, p^d)}{\partial^d x^{dk}},
%(A.7)
\label{eq:a-7}\end{equation}
 or in the unified notation

\begin{equation}
\omega^d_{\mu\nu}\times^d\frac{d^d  b^{d\nu}}{d^d
  t^d}\;=\;\frac{\partial^dH^d(t^d,b^d)}
 {\partial^db^{d\mu}},
%(A.8)
\label{eq:a-8}\end{equation}
where $\omega^d_{\mu\nu}$ is  the  {\it isodual  canonical  symplectic
tensor}

\begin{equation}
(\omega^d_{\mu\nu})\;=\;(\partial^dR^{\circ
d}_\nu/^d\partial^db^{d\mu}            -            \partial^dR^{\circ
d}_\mu/^d\partial^db^{d\nu})\;=\; \left(\begin{array}{cc}  0&I\\  -I&0
\end{array}\right) \;=\; -(\omega_{\mu\nu}).
%A.9
\label{eq:a-9}\end{equation}

Note that in  matrix form  the  canonical symplectic tensor is  mapped
into the canonical Lie tensor.

       The isodual {\it Hamilton--Jacobi equations} are then given by

\begin{equation}
\partial^d    A^{\circ    d}/^d\partial^dt^d  +     H^d  \;=\;   0 ,\;
\partial^dA^{\circ  d} /^d  \partial^d x^d_k  - \hat{p}_k  \;=\; 0 ,\;
\partial^dA^{\circ d} /^d \partial^dp^d_k \equiv 0 .
%(A.10)
\label{eq:a-10}\end{equation}

        The  {\it isodual Lie  brackets}  among two isodual  functions
$A^d$ and $B^d$ on $S^d(t^d,x^d,p^d)$ then become

\begin{equation}
[A^d,B^d]^d\;=\;\frac{\partial^dA^d}{\partial^db^{d\mu}}d\;\times^d
\omega^{d\mu\nu}\times^d
\frac{\partial^dB^d}{\partial^db^{d\nu}}d\;=\;-[A,B]
%(A.11)
\label{eq:a-11}\end{equation}
where
\begin{equation}
\omega^{d\mu\nu} \;=\; [ ( \omega^d_{\alpha\beta} )^-1]^{\mu\nu} ,
%   (A.12)
\label{eq:a-12}\end{equation}
is  the  {\it isodual Lie  tensor}. The   direct representation of the
isodual Newton equations in first--order form is self--evident.

        In  summary, all  properties of   the  isodual  theory at  the
Newtonian  level    carry over  at  the level   of isodual Hamiltonian
mechanics. In so doing, there is the emergence of a fundamental notion
of these studies, the  characterization of antimatter via {\it isodual
space--time   symmetries},    i.e.,   the  isodual  Galilean  symmetry
$G^d(3.1)$  for  nonrelativistic treatments,  the isodual Poincar\`{e}
symmetry  $P^d(3.1)$  for  relativistic treatments   and  the  isodual
isoPoincar\`{e} symmetry for gravitational treatments~[6].

\vskip 0.5cm
         {\bf Isodual naive quantization}.
The isodual   Hamiltonian mechanics and  its underlying
isodual symplectic geometry permit the   identification of the   novel
{\it naive isodual quantization}

\begin{eqnarray}
A^{\circ   d}&   \rightarrow&  -    i^d \times^d     \hbar^d  \times^d
Ln^d\psi^d(t^d, r^d), \\
%  (A.13) \label{eq:a-13}
\partial^d  A^{\circ d}/^d\partial^dt^d +  H^d \;=\;  0 & \rightarrow&
i^d \times^d \partial^d \psi^d  /^d \partial^d t^d \;=\; H^d \times^d
\psi^d \;=\; E^d \times^d \psi^d , \\
%(A.14)
\partial^dA^{\circ   d}  /^d \partial^d  x^{dk}   -  \hat{p}_k \;=\; 0
&\rightarrow& p_k^d \times^d \psi^d   \;=\; - i^d \times^d  \partial^d
\psi^d , \; \\ %(A.15)
\label{eq:a-15}\end{eqnarray}
or more refined   isodualities of symplectic quantization  (see, e.g.,
Ref.~[24] for the conventional case),
which characterize a novel  image of quantum mechanics for
antiparticles, called {\it isodual quantum mechanics},
studied in the joint paper~[10].

\vspace*{2cm} \centerline{ \large \bf Acknowledgements.}

        The  author  would like  to  express  his  appreciation to all
participants of the {\it  International Workshop on Antimatter Gravity
and  Anti-Hydrogen  Atom Spectroscopy}  held   in Sepino,
Molise, Italy,  in May 1996, for invaluable  critical
comments.     Particular   thanks   are    also    due  to  Professors
A. K.  Aringazin,  P.  Bandyopadhyay,   S. Kalla,  J. V.   Kadeisvili,
N.  Kamiya, A.  U. Klimyk,  R.  Miron, R.  Oehmke, G.   Sardanashvily,
H. M.   Srivastava, T. Gill,  Gr.  Tsagas, N.  Tsagas, C.  Udriste and
others for  penetrating comments.  Special  thanks are finally due  to
M. Holzscheiter and J. P.  Mills, jr., for invaluable critical remarks
and  to H.  E.  Wilhelm  for  a   detailed  critical reading of   this
manuscript.


\newpage
\begin{thebibliography}{00}
\bibitem{r1}  P.   A.  M. Dirac,   {\it   The  Principles  of  Quantum
   Mechanics}, Clarendon Press, Oxford, fourth edition (1958).

\bibitem{r2} R.   L.   Forward,  in {\it  Antiprotons   Science  and
  Technology}, B.   W.   Augenstein, B. E.  Bonner,  F.  E. Mills  and
  M. M. Nieto, Editors, World Scientific, Singapore (1988)

\bibitem{r3} R. M. Santilli,  Hadronic J. {\bf \underline{8}}, 25  and
     36 (1985)

\bibitem{r4} R. M.   Santilli,  Algebras, Groups  and Geometries  {\bf
   \underline{10}}, 273 (1993)

\bibitem{r5.}  R. M. Santilli, Comm. Theor. Phys. {\bf \underline{3}},
153(1993)

\bibitem{r6.}    R.  M.    Santilli,   J.  Moscow    Phys.  Soc.  {\bf
\underline{3}}, 255 (1993)

\bibitem{r7.}  R. M.  Santilli, Hadronic J.  {\bf \underline{17}}, 257
(1994)

\bibitem{r8.}  R.  M.  Santilli,   Rendiconti  Circolo  Matematico  di
Palermo, Supplemento {\bf \underline{42}}, 7 (1996)

\bibitem{r9.}  R. M. Santilli,  {\it Elements of  Hadronic Mechanics},
Vol.  II: {\it Theoretical  Foundations}, Ukraine Academy of Sciences,
Kiev, Second Edition (1995)

\bibitem{r10.}  R. M. Santilli,  Does  antimatter emit  a new light  ?
Hyperfine Interactions, in press.

\bibitem{r11.}  R. M.  Santilli,  Relativistic hadronic  mechanics:  A
nonunitary,  axiom--preserving  completion  of  relativistic   quantum
mechanics, in press at a leading physics journal.

\bibitem{r12.}  J. P. Mills,  jr, Hadronic J. {\bf \underline{19}},  1
(1996)

\bibitem{r13.}  J. V. Kadeisvili, Algebras, Groups and Geometries {\bf
\underline{9}}, 283 and 319 (1992);  Math. Methods in Appl. Sci.  {\bf
\underline{19}},   1349   (1996);  {\it     Santilli's   Isotopies  of
Contemporary Algebras,  Geometries and Relativities},
Second Edition,  Ukraine Academy  of Sciences,  Kiev  , in
print

\bibitem{r14.}  J. L\^{o}hmus, E.  Paal   and L. Sorgsepp,    {\it
Nonassociative Algebras in  Physics}, Hadronic Press, Palm Harbor, FL,
USA  (1994).  D. S.   Sourlas  and  G.  T. Tsagas,  {\it  Mathematical
Foundations of the Lie-Santilli  Theory}, Ukraine Academy of Sciences,
Kiev (1993)

\bibitem{r15.}   M. Holzscheiter,  Editor,   {\it Proceedings   of the
International Workshop  on Antimatter Gravity}, Sepino, Molise, Italy,
May 1996, Hyperfine Interactions, in press

\bibitem{r16.}  E. C.  G.   Sudarshan and N. Mukunda,  {\it  Classical
Mechanics: A Modern Perspective}, Wiley \& Sons, New York (1974)

\bibitem{r17.}  W. Pauli, {\it Theory  of Relativity}, Pergamon Press,
London (1958)

\bibitem{r18.}   C.   W. Mismer,  K. S.   Thorne and  A. Wheeler, {\it
Gravitation}, Freeman, San Francisco (1970)

\bibitem{r19.}  P. Freud,   Ann. Math. {\bf  \underline{40}} (2),  417
(1939)

\bibitem{r20.}   W.     M.  Fairbank       and  F.  C.      Witteborn,
Phys. Rev. Lett. {\bf \underline{19}}, 1049 (1967)

\bibitem{r21.}    R.  E.  Brown    et   al.,  Nucl. Instr.     Methods
Phys. Res. {\bf \underline{B56}}, 480 (1991)

\bibitem{r22.}  M.   M. Nieto and   T.  Goldman, Phys.   Reports  {\bf
\underline{205,}}221 (1991), erratum {\bf \underline{216}}, 343 (1992)

\bibitem{r23.}  R. M.  Santilli, Ann.  Phys. {\bf \underline{83}}, 108
(1974)

\bibitem{r24.}  J. Sniatycki, {\it  Geometric Quantization and Quantum
Mechanics}, Springer--Verlag, New York (1979)
\end{thebibliography}
\end{document}